\documentclass[prd, aps, superscriptaddress, preprintnumbers, twocolumn, floatfix, nofootinbib]{revtex4}

\usepackage{amsmath}    
\usepackage{graphicx}   
\usepackage{verbatim}   
\usepackage{color}      
\usepackage{subfigure}  
\usepackage{hyperref}   
\usepackage{float}     
\usepackage{mdwlist}

\def\be{\begin{equation}}
\def\ee{\end{equation}}

\begin{document}

\title{Signature of a Cosmic String Wake at $z=3$}

\author{Disrael Camargo Neves da Cunha\footnote{camargod@hep.physics.mcgill.ca}}
\affiliation{Department of Physics, McGill University,Montreal, QC, H3A 2T8, Canada and \\
Cosmology, Universe and Relativity at Louvain,
Institute of Mathematics and Physics, Louvain University, 2 Chemin
du Cyclotron, 1348 Louvain-la-Neuve, Belgium }


\date{\today}

\begin{abstract}
In this paper, we describe the results of N-body simulation runs, which include a cosmic string wake of tension $G\mu= 4 \times 10^{-8}$ on top of the usual $\Lambda CDM$ fluctuations. To obtain a higher resolution of the wake in the simulations compared to previous work, we insert the effects of the string wake at a lower redshift and perform the simulations in a smaller volume. A curvelet analysis of the wake and no-wake maps is applied, indicating that the presence of a wake can be extracted at a three-sigma confidence level from maps of the two-dimensional dark matter projection down to a redshift of $z=3$.

\end{abstract}

\maketitle

\tableofcontents

\section{Introduction}

Cosmic strings are linear topological defects in Quantum Field Theory, which exist as solutions in some models that go beyond the Standard Model of Particle Physics \cite{CosmicStrings}. A cosmic string consists of a one-dimensional region of trapped energy, having significant gravitational effects for cosmology. If a model of nature admits cosmic string solutions, strings will necessarily form during the early universe \cite{Kibble}. For example, in some models, they form after the end of inflation, and in others, they form during a phase transition in the early radiation phase of Standard Big Bang Cosmology. After the cosmic strings form, they will persist as a scaling network. This means that the network of cosmic strings will have the same properties at all times if we scale the length observables to the Hubble radius \cite{Kibble}. The network will consist of a few long strings moving near the speed of light and also of loops of different sizes, and it will source sub-dominant fluctuations at all times. 
The gravitational effects of a cosmic string are characterized by only one number $\mu$, its tension, which does not affect the scaling solution properties of the cosmic string network. The tension can also be seen as the energy per unit of length of the cosmic string, and it is related to the energy scale $\eta$ at which the strings form by the following equation:
\be
G\mu \simeq {(\eta/m_{pl})}^{2}
\ee
where $G$ is Newton's constant and $m_{pl}$ is the Planck mass. 
The presence of cosmic strings does not produce acoustic oscillation features on the Cosmic Microwave Background (CMB) angular power spectrum. This fact contributes to the current upper bound \footnote{Note that there are stronger limits on the
string tension which comes from limits on
the stochastic background of gravitational
waves on length scales which the pulsar timing
arrays are sensitive to (see e. g. \cite{Arzoumanian:2018saf}).
These bounds come from gravitational
radiation from string loops, and assume a
scaling distribution of string loops where the
total energy in strings is dominated by the
loops \cite{CSsimuls}.
However, field theory cosmic string simulations
\cite{small} do not yield a significant
distribution of string loops. Thus, bounds
on the cosmic string tension from gravitational
radiation from string loops are less robust than
the ones coming from the long strings.} on the cosmic string tension \cite{Dvorkin:2011aj}:
\be
G\mu \approx 1.5 \times 10^{-7}
\ee

A good study on the observational aspects of cosmic strings has two possible outputs \cite{RHBtopRev}. One possibility is observing a cosmic string, which would be a significant achievement on probing particle physics models beyond the Standard Model of Particle Physics. The other option is not to observe cosmic strings, which will lower the bound on the cosmic string tension, thus ruling out classes of particle physics models. Besides this, cosmic strings could produce interesting results for cosmology, such as explaining the origin of Fast Radio Bursts \cite{Brandenberger:2017uwo}, primordial magnetic fields \cite{Brandenberger:2009ia}, and the origin of supermassive black holes \cite{Bramberger:2015kua}. 

This work will concentrate on the Large Scale Structure (LSS) as a complementary (in addition to the CMB) arena for probing cosmic string. The primary motivation is that LSS data contains three-dimensional information, which includes many more modes than the two-dimensional maps from the CMB. The disadvantage of LSS is that the effects of non-linearities are essential, so theoretical predictions are harder to be obtained. This study concentrates on one of the main effects of cosmic strings for LSS, which is the cosmic string wake. A cosmic string wake is a planar overdense region that forms behind a long string as it passes by the matter distribution \cite{wake}. This effect is a consequence of the fact that a space perpendicular to the long string will have a missing angle given by $\alpha=8\pi G\mu$ \cite{CosmicStrings}, causing dark matter particles to receive a velocity kick towards behind the string as soon as it passes by between them. The expression for the velocity perturbation is the following:
\be
\delta v = 4\pi \gamma_{s} v_{s} G\mu
\ee
where $v_{s}$ is the transverse velocity of the string, and  $\gamma_{s}$ is the associated Lorentz factor. The velocity kick makes the particles meet behind the string, forming a wedge-like structure with two times the average matter density. This is the wake. The initial geometry of the wake after formation at $t=t_{wf}$ will be a box of volume $V$, consisting of two large planar dimensions of the order of the Hubble radius $\approx t_{f}$ and one smaller thickness with a length of the order of the Hubble radius multiplied by the deficit angle:
\be
V\approx t_{f} \times t_{f} v_{s} \gamma_{s} \times 4 \pi G\mu t_{f} v_{s} \gamma_{s} \label{wake_volume}
\ee

At early times it is possible to obtain an analytical understanding of the wake evolution thanks to the fact that the matter fluctuation outside the wake was in the linear regime. The Zeldovich approximation \cite{Zeld} gives the evolution of the comoving wake thickness $\psi_{3}$ as a function of redshift $z$ \cite{wakegrowth} :
\be
\psi_{3}=\frac{24\pi}{5}G\mu v_{s}\gamma_{s}t_{0}\frac{\sqrt{1+z_{eq}}}{(1+z)}
\ee
where $t_{0}$ is the present time, and $z_{eq}$ is the redshift of matter and radiation equality. Note that the wake produces a nonlinear density fluctuation at arbitrarily early times. Since structures start to grow only after the time of equal matter and radiation, we choose this time as the time for wake formation (so we will consider $t_{f}=t_{eq}$). As the thickness grows as in linear theory, the planar dimension increases just with the Hubble flow, and are fixed in comoving coordinates. For the formation time we are considering, the planar dimension of the wake is about $\approx 100Mpc/h$. If the analytical thickness evolution remains valid up to today, we would have wakes with the thickness of $\approx 0.08Mpc/h$ at present (for $G\mu=4 \times 10^{-8}$ and $v_{s}\gamma_{s} \approx 0.4$). 

Once the $\Lambda CDM$ perturbations enter the nonlinear regime (at about the time of re-ionization), the local mass distribution in the wake becomes highly nonlinear, the $\Lambda CDM$ fluctuations will disrupt the wake \cite{cunha16}, and the subsequent dynamics has to be studied numerically. In this paper, we simulate cosmic string wakes using an N-body code called CUBEP3M \cite{code} and apply a statistic that extracts the wake signal. 

In a recent paper \cite{cunha18} (see also \cite{Laliberte:2018ina}) we began a study of the dark matter distribution induced by string wakes \cite{wake} inserted at $z=31$, using a simulation box of lateral size $L=64Mpc/h$ and $np=512$ particles per dimension. We found that the string signals for a wake with a string tension of $G\mu=10^{-7}$ can be identified down to a redshift of $z\geq 10$. A possible cause for not being able to identify the wake at lower redshifts comes from the fact that the wake thickness was about one order of magnitude smaller than the resolution length of the simulation grid at the time of wake insertion. The fact that the wake survives down to redshift $z=10$ supports the idea that the wake global signal remains present despite losing its local signal \cite{cunha16}. In the current work, we take a complementary approach, and we use a box with a small lateral size (of $L=4Mpc/h$) and a higher number of particles ($np=1024$ particles per dimension ), so the wake becomes well-resolved at the time of insertion, even for a wake of smaller tension ($G\mu=4 \times 10^{-8}$ is the smallest tension for the wake to be well resolved with the specifications given above).  The downside of this approach is that we lose part of the global wake signal, and the advantage is that the effect of the wake lasts for more time.

Ultimately we would like to obtain an experimental prediction for the observability of the wake (having a $G\mu=4 \times 10^{-8}$ as the tension for the cosmic string that originates it). The present work has the limitation that it is valid only for dark matter maps. Nevertheless, having a method for wake detection in pure dark matter maps gives the hope that it can be used (or extended) to baryonic maps, since baryons trace dark matter.  It will be explored in future work if this tracer is enough to maintain the wake signal. Assuming that this is possible, one can ask which experiments will be able to extract a cosmic string wake signal. In that respect, at the redshifts of interest ($z = 3$), the wake thickness would be well resolved for an angular resolution of $\delta \theta \approx 0.9\ arcsec$ and a redshift resolution of $\delta z \approx 2.8\times 10^{-5}$. From the near future experiments, SKA-mid \cite{SKA} will be able to provide the required angular resolution (about two times finer), but its redshift resolution will be about six times coarser. Therefore we will assume during the analysis that the angular resolution is high enough to resolve the wake, but we will also consider a coarse redshift resolution (more than six times coarser than the wake thickness). 

The paper is organized as follow: section II contains a discussion on the previous works regarding wake evolution in the nonlinear regime; Section III describes the simulations performed in the present work, and the analysis of the data described in section III is performed in section IV; Finally, in section V we summarize the essential results obtained and indicate possible paths for an extension of the analysis.

\section{Review of Cosmic string wakes in the non-linear regime}

The wake produces a planar non-linear density perturbation at arbitrarily early times, so early on, the wake is unambiguously distinguishable from $\Lambda CDM$ fluctuations. Once the $\Lambda CDM$ fluctuations start to dominate, nearby halos begin to accrete material from the wake, causing wake fragmentation.
An analytical study regarding the wake disruption by $\Lambda CDM$ fluctuations was presented in \cite{cunha16}, where two criteria for wake disruption were introduced. The first one concerns local stability, which was studied by considering a cubic box with the dimension of the wake thickness and computing the standard deviation of the density contrast in this region from $\Lambda CDM$ fluctuations. The second criterion takes the global extension of the wake region into account, by computing the standard deviation of the density contrast from $\Lambda CDM$ inside a box with dimension $V$ (see (\ref{wake_volume})) given by the whole wake. Both conditions were computed, and the main result indicates that although a $G\mu=4\ \times 10^{-8}$ wake could be locally disrupted at $z\approx 8$, it could, in principle, be distinguished from $\Lambda CDM$ fluctuations at all times using the global information of the wake.

A method used to extract the wake signal from the dark matter distribution was presented in \cite{cunha18} and can be summarized as follows: for any direction of the sphere, we consider an associated projection axis passing through the origin of the box. We then consider slices of the simulation box perpendicular to that axis at each point $x$ of this axis with thickness given by the grid size of the simulation, and we compute the mass density $\delta (x)$ of dark matter particles in that slice as a function of $x$. A one-dimensional filter wavelet analysis is then performed on the mass density $\delta (x)$, giving a filtered version of it, called $f\delta$. We then compute the maximum value S of $f\delta (x)$ for each direction (pair of spherical angles) in the simulation box. S is a map on the surface of the sphere (which represents the spherical angles), and $\hat{S}$ is its maximum value. The signal to noise ratio for the spherical peak($max(S)$) divided by standard deviation ($max(S)/std(S)$) distribution for ten simulations without wake and three simulations with a $G\mu=10^{-7}$ wake was found to be ${\cal \bar{R}}=8.1$ at redshift $z=10$ and insignificant at lower $z$.

The setup of the N-body simulation is the following. We used an N-body simulation program called {\small CUBEP$^3$M}, a public high performance 
cosmological $N$-body code based on a two-level mesh gravity solver augmented with sub-grid particle-particle interactions \cite{code}. This code generates and evolves initial conditions (which are realizations of $\Lambda CDM$ fluctuations) containing positions and velocities of particles inside a cubic box. There is an option to save the phase space of the distribution at any redshift and rerun the code from this checkpoint. We use this feature for wake insertion, by modifying the saved phase space distribution at time $t_i$ by including effects of a wake. The modification consists in displacing and giving a velocity kick to particles towards the wake plane. The absolute value of the displacement $\psi(t_i)$ and velocity perturbation $\dot{\psi}(t_i)$ in comoving units can be computed using the following equations (see \cite{wakegrowth} for details)
\be
\psi(t_i) \, = \, \frac{3}{5} 4 \pi G\mu v_s \gamma_s t_{eq} z(t_{eq}) \frac{z(t_{eq})}{z(t_i)} \, .
\label{displacement}
\ee
and
\be
{\dot{\psi}}(t_i) \, = \, \frac{2}{5} 4 \pi G\mu v_s \gamma_s t_{eq} z(t_{eq}) \frac{z(t_{eq})}{z(t_i)} \frac{1}{t_i} \, .
\label{velocity}
\ee
Once the wake insertion is made, the new modified distribution is further evolved by the N-body code. 

%
%
%
%
%
%
%
%

The primary goal of this paper is to find a statistic that can extract the wake presence without using information about the simulation without a wake. In the next section, we will describe the simulations in more detail.

\section{Simulations}

We performed ten N-body simulations without wakes with the following cosmological parameters: $\Omega_{\Lambda}=0.7095$, $\Omega_{\rm b}=0.0445$, $\Omega_{\rm CDM}=0.246$, $n_{t}=1$, $n_{s}=0.96$, $\sigma_{8}=0.8628$, $h=0.70$, $T_{\rm cmb}(t_{0})=2.7255$, $z_{eq}=3400$. Those simulations consists of $np=1024$ particles per dimension, a lateral size of $L=4Mpc/h$, and initial conditions generated at $z=31$, with checkpoints at $z=15$, $z=10$. $z=7$, $z=5$, $z=3$ and $z=2.9$. In each simulation above the wake (with $G\mu = 4 \times 10^{-8}$) is inserted at the $z=10$ checkpoint and further evolved, having the same remaining redshifts. All simulations used 512 cores divided into 64 MPI tasks and were run in the Niagara cluster, of SciNet, part of the Compute Canada consortium.

The reason the wake is introduced at $z=10$ is due to the requirement that the wake should be well resolved. In this specification, the wake thickness contains about one simulation cell at the insertion. An earlier insertion (and therefore a thinner wake thickness) would require a simulation with more than 1024 particles per dimension for it to be well resolved, which, although possible, is computationally costly. A late insertion becomes dangerous if the wake is locally disrupted. In this case, there is no reason to believe that the analytical prediction for the local wake density and thickness would work. Here there is not such a problem since the wake is expected to become locally disrupted only at $z \approx 8$ when the scale of the wake thickness is expected to become non-linear \cite{cunha16}.

As pointed out in \cite{cunha18}, if we displace all particles towards a central plane, this creates a nonphysical void on the parallel planes at the boundary of the simulation box. Here we circumvent this problem by using a suppression of the velocity and displacement perturbations that starts halfway between the wake and the boundary and linearly decreases to zero at the boundary. This procedure avoids the creation of a planar void at the boundary since the particles are not displaced there.

\begin{figure}[h]
\centering

\includegraphics[width=.45\textwidth]{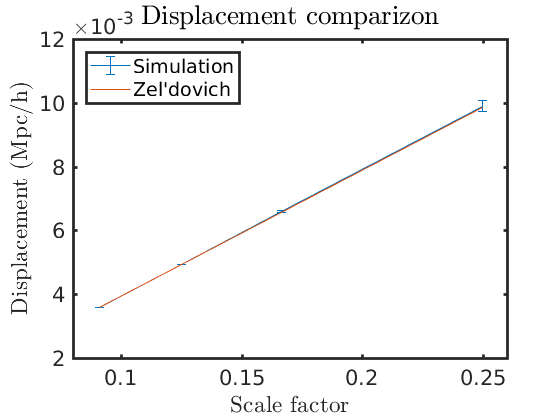}

\caption{Average displacement induced by the wake (in blue), compared with the analytical prediction (in red). The error bars are the standard deviation computed from 10 simulations }
\label{check_zel}

\end{figure}

Figure \ref{check_zel} shows the average displacement induced by the wake on each particle and compares it with the analytical prediction, showing a good agreement between the two.

\begin{figure}[h]
\centering 

\includegraphics[width=.45\textwidth]{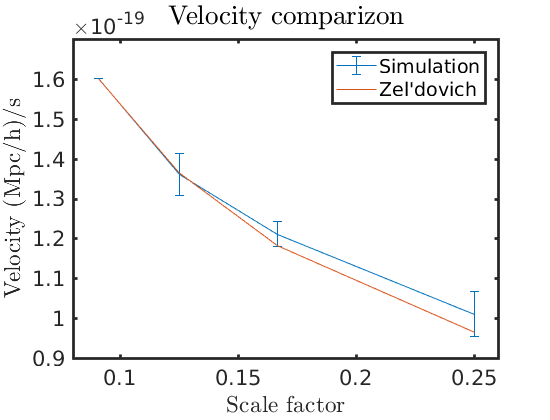}

\caption{Average velocity perturbation induced by the wake (in blue), compared with the analytical prediction (in red). The error bars are the standard deviation computed from 10 simulations }
\label{check_vel}

\end{figure}

A similar analysis was done for the induced velocity perturbation. As figure \ref{check_vel} shows, the difference between the numerical and analytical velocity perturbations is not significant.

\section{Analysis}

\begin{figure}[h]
\centering

\includegraphics[width=.45\textwidth]{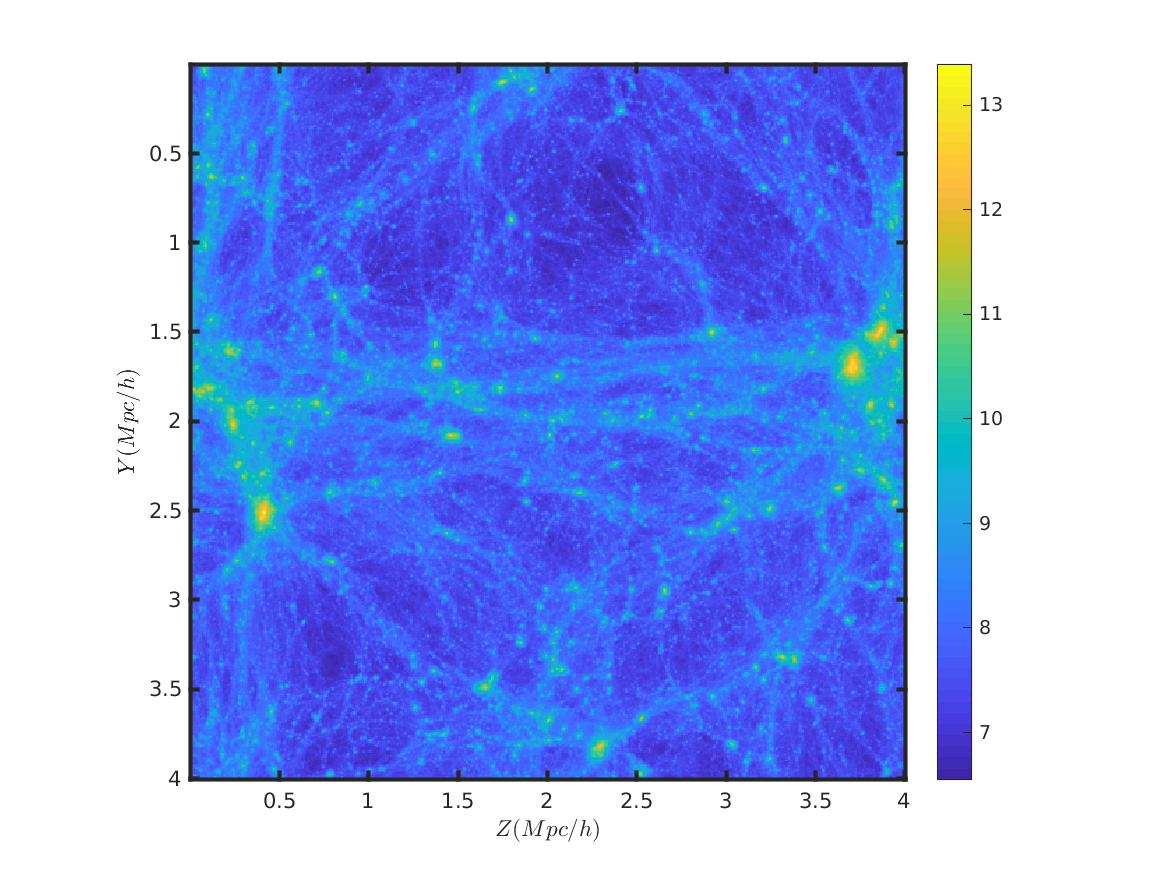}

\includegraphics[width=.45\textwidth]{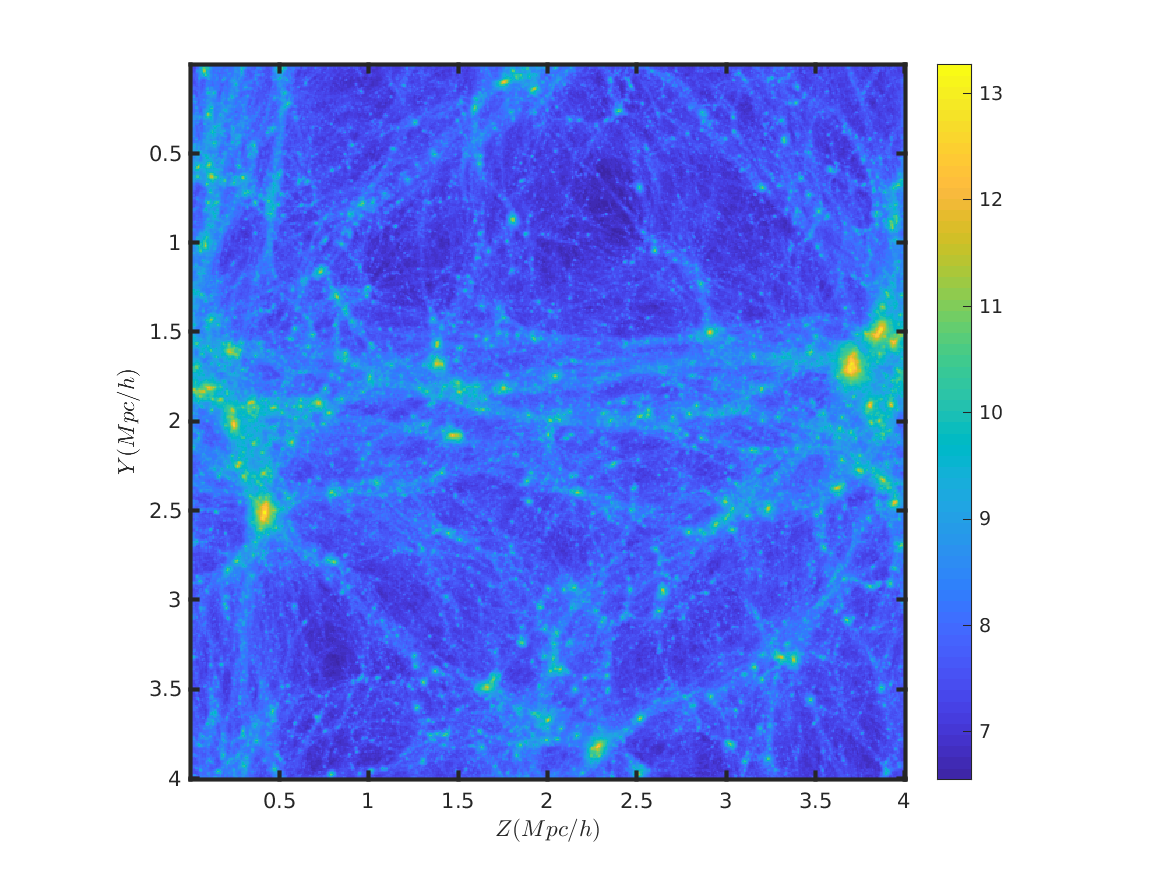}

\caption{Logarithm of the mass of a two-dimensional projection of simulation particles for lateral size  $L=4Mpc/h$ and redshift $z=3$. The upper plot shows the simulation without the wake, and the plot on the bottom shows the simulation with a $G\mu=10^{-7}$ wake at the central position of the $z$ axis.}
\label{2d_proj_z3s10}

\end{figure}

The wake presence is not clear if we compare the visualization of a two-dimensional projection for a pure $\Lambda CDM$ simulation with a $\Lambda CDM$ plus wake simulation. Figure \ref{2d_proj_z3s10} illustrates this at $z=3$. Both figures show the logarithm of two-dimensional projections of the dark matter distribution (in simulation units). The upper figure contains just $\Lambda CDM$ fluctuations, and the bottom figure contains the same $\Lambda CDM$ fluctuations plus a $G\mu=4\times 10^{-8}$ wake located at the plane $Z\approx 2Mpc/h$ which under projection would appear as a vertical line at the middle of the panel (since we are projecting onto a plane perpendicular to the wake plane). Both figures are almost similar, and the wake presence is not clear by eye. It was not possible to find a good statistics that extracts the wake signal for such projections.


\begin{figure}[h]
\centering

\includegraphics[width=.45\textwidth]{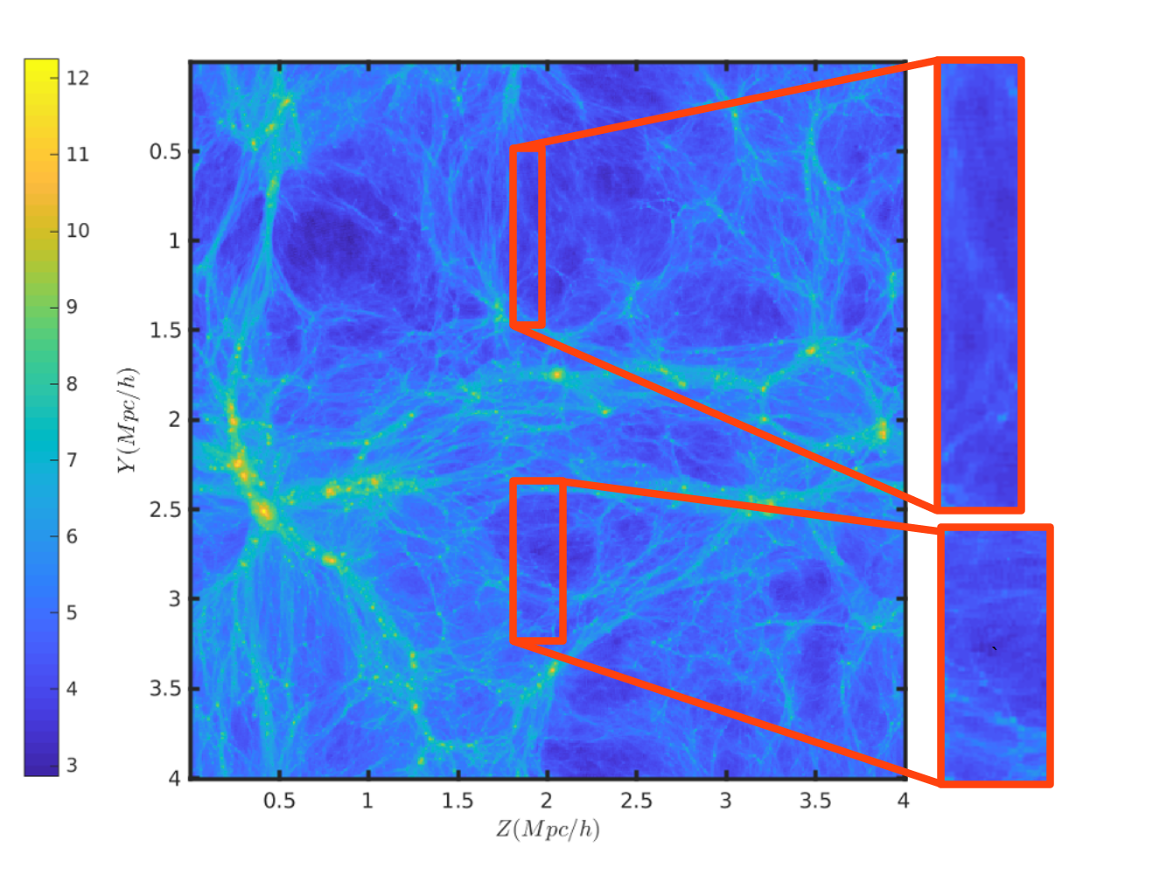}
\includegraphics[width=.45\textwidth]{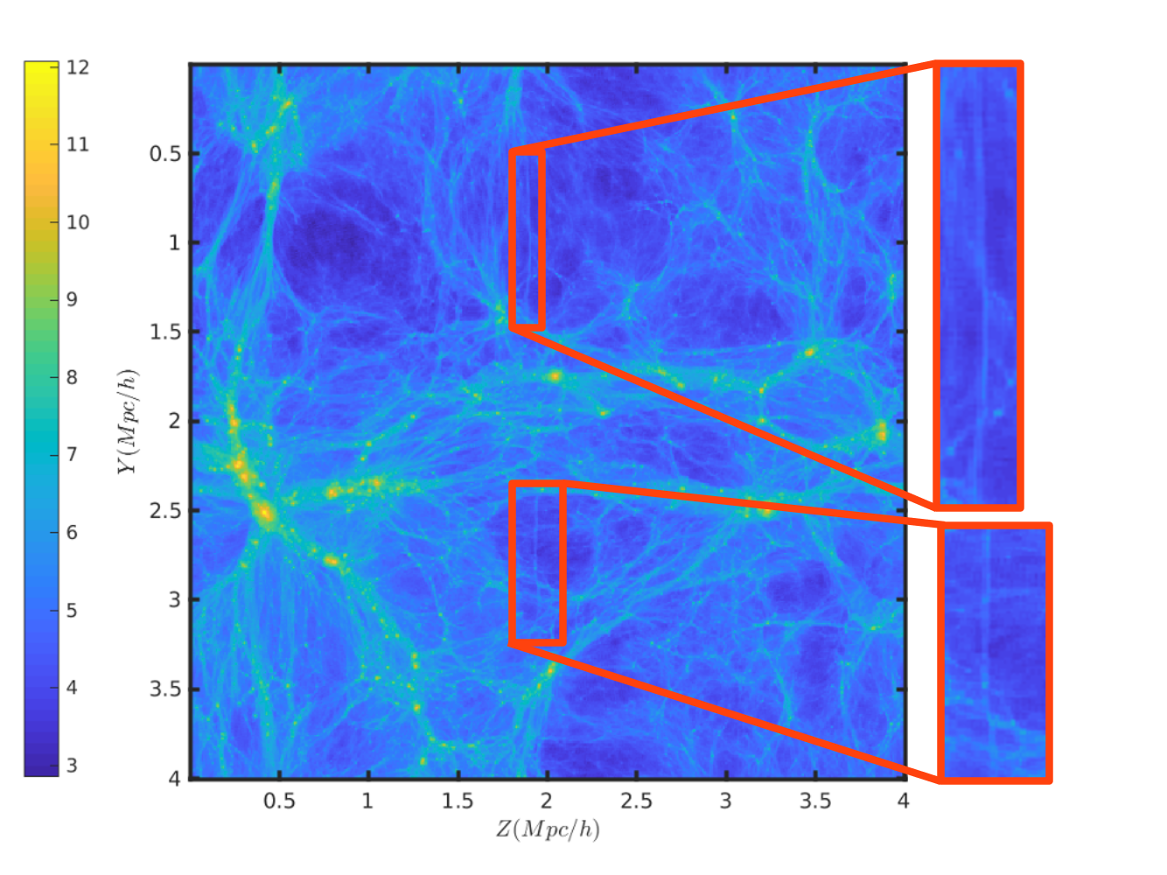}

\caption{Logarithm of the mass of a two-dimensional projection of a slice of the simulation box with lateral size $L=4Mpc/h$ and redshift $z=3$. The upper plot shows the slice without the wake, and the plot on the bottom shows the slice with a $G\mu=4 \times 10^{-8}$ wake. Two red rectangles at the central position of the $Z$ axis are made, which was augmented so the wake can be better visualized in comparison with the simulation without a wake.}
\label{2d_proj_z3s10_sl31}

\end{figure}

This occurs because the Gaussian fluctuations displace the particles on the wake, and these particles no longer form a straight plane.  However, the wake can be better recovered if we project not the entire $X$ direction, but several slices of it. From an experimental point of view, the location of the pixels in the map ($Y$ and $Z$ direction) corresponds to angular coordinates, and the depth ($X$ coordinate) corresponds to redshift. Taking SKA as an example \cite{SKA}, it would allow us to perform 32 slices on the simulation volume long the $X$ direction, and for each slice to produce a figure with $512\times 512$ pixels of resolution. It was found that if we slice the $X$ direction in 8 different parts and perform a two-dimensional projection on each slice, the wake can be better visualized. Figure \ref{2d_proj_z3s10_sl31} shows the slice number 6 associated with the same simulations of figure \ref{2d_proj_z3s10}. The wake presence is more explicit in this case and corresponds to a vertical line at the middle of the panel on the bottom figure of \ref{2d_proj_z3s10_sl31}, in which a zoom on the wake is made to facilitate visualization.


%

A visual identification of the wake does not mean that a wake detection is granted since the wake presence should be obtained quantitatively and without previous knowledge of the simulation without a wake. A pipeline for a quantitative measure of the wake presence will be developed in the next subsection. The initial input will consist of a three-dimensional dark matter map of ${(4Mpc/h)}^3$ (in comoving coordinates) at $z=3$. We choose to use a resolution of this map to be the same as provided by $SKA$ \cite{SKA}: $512\times 512 \times 32$, with the first two dimensions corresponding to the angular coordinates and the third one corresponding to redshift coordinate. The output of the pipeline will be a number $S$ that is designed to contain the information about the wake presence.

In subsection B, we will show the result of a statistic that analyses a set of two-dimensional projections of the dark matter particles. We ask the question if it is possible to differentiate the three-dimensional map with its third-dimensional axis perpendicular to the wake from other projections without the wake.\footnote{ One motivation for this approach is that some experiments, like SKA \cite{SKA}, have a poor redshift resolution compared with the angular resolution, so the data set can be better viewed as layers of two-dimensional intensity maps with each layer corresponding to a redshift bin.}

A sky map with a specific redshift layer can be further subdivided into several square images (using small angle approximation), and in a universe with a cosmic string network, there is a small non-zero probability that one of those images has a wake perpendicular to it. We will see in subsection C that this small fraction is still sufficient to pinpoint a universe with $G\mu = 4 \times 10^{-8}$ cosmic strings at redshift down to $z=3$, without assuming any information on the wake orientation. The idea of the analysis presented in subsection C is that the signal $S$ (described in subsection A) will have larger values in maps with wakes in the optimal orientation. We can compute the distribution of $S$ in maps with and without wakes without assuming any orientation of the wake and declare that above a specific threshold $S_{th}$, we have a wake detection. We call such a sample with $S>S_{th}$ an outlier. From the simulations, we will obtain the probability detecting an outlier in the two different cases: with wakes and in maps without wakes. Once those probabilities are computed, we will ask how many three-dimensional maps, similar to the ones used in the analysis will be provided by an experiment ($SKA$), giving us the total number of three-dimensional maps that will be available to probe. Finally, we will obtain the probability of having more than a specific number of outliers from all the available maps, both in a universe with wakes and in a universe without wakes.

All the analyses were performed in the Graham Cluster of Calcul Quebec, part of the Compute Canada consortium.



\subsection{A curvelet filtering of the two-dimensional projections} \label{sec_3a}

Here we describe the pipeline for the statistics that we use for wake detection. All computations were performed using Matlab, together with the curvelet package CURVELAB \cite{dcurvelet}. We will also use ridgelet transformation \cite{radon}, which detects straight lines (ridges) that cross an entire image. Similarly, the curvelet base functions have line segments domains, which can be seen as a local version of the ridges. 


The first of the filtering procedure step is, for each cubic simulation box, to slice it into 32 different tiles \footnote{In the optimal case the slice is perpendicular to the $X$ direction, while the wake is parallel to it} and to obtain a two-dimensional map of it by projecting the associated slice onto the $Y-Z$ plane, forming 32 images of $512 \times 512$ pixels. 

For each two-dimensional dark matter map particle number (which is proportional to the dark matter mass) $pn$ (viewed as a two-dimensional array with the projected number of particles as each one of its elements) we perform the following steps: 

1-) Compute the arc-tangent transformation $at=arctg(16*(1+dc))$ of the density contrast $dc$ of the dark matter, where $dc=(pn-\bar{pn})/\bar{pn}$, and $\bar{pn}$ is the average dark matter for the whole simulation. The number one is added on the density contrast for the computation of the arc-tangent to the result to be strictly positive. The number $16$ is multiplied in order to bring regions with high-density contrast closer to the regions with lower density contrasts.

2-) Compute the curvelet-filtered transformation $curv(s,w,i,j)=C(at)$, were $s$ corresponds to the scale of the ridge, $w$ to the angle, $i$ and $j$ to the positions. All the curvelet coefficients with scales higher than the expected wake thickness were set to zero.

3-) A wake ridge corresponds to coefficients that are much higher than the others if we fix the position, scale and allow the angles to vary. Therefore we would like to highlight the maximum of the function $f_{s,i,j}(w)=curv(s,w,i,j)$ in the angle variable $w$. We implement that for each $(s,i,j)$ by multiplying the function $g(w)=f_{s,i,j}(w)$ by its own density contrast $\tilde{g}(w)=(g(w)-\bar{g})/\bar{g}$, where $\bar{g}=mean(g(w))$. The resulting new wavelet coefficient becomes $\tilde{f}_{s,i,j}(w)=f_{s,i,j}(w)\times \tilde{g}(w) $.

4-) After the previous filter, the inverse curvelet transformation is taken and $dm_{2d}=curv^{-1}(\tilde{f}_{s,i,j}(w))$. 

Once the 32 filtered images are computed, we combine them together again in a three-dimensional map $dm_{3d}$, where one of the dimensions corresponds to the direction of slicing (and go from one to 32) and the remaining two dimensions are the labels for the filtered image pixels. The image of each slice corresponds to the filtered image obtained in the steps above for each one of the slices.

5-) Compute a 3d curvelet decomposition $curv3d=C(dm_{3d})$, and aply a generalization of the filter presented in step 3, but this time there are two labels for the angles and three labels for the position of the local three dimensional ridge (a planar segment). By doing the inverse curvelet transformation, we obtain 32 filtered slices $dmFilt(i,j,k)=C^{-1}(curv3d)$ from the original 32 slices, where $i$ and $j$ range from $1$ to $512$ and $k$ range from $1$ to $32$.

6-) We add the values of the pixels for four adjacent slices (in the $k$ direction). For each $i,j$ pixel, we add the values of four consecutive pixels in the $k$ direction. This results in 8 slices, where $i$ and $j$ range from $1$ to $512$ and $k$ range from $1$ to $8$.

\begin{figure}[h]
\centering

\includegraphics[width=.45\textwidth]{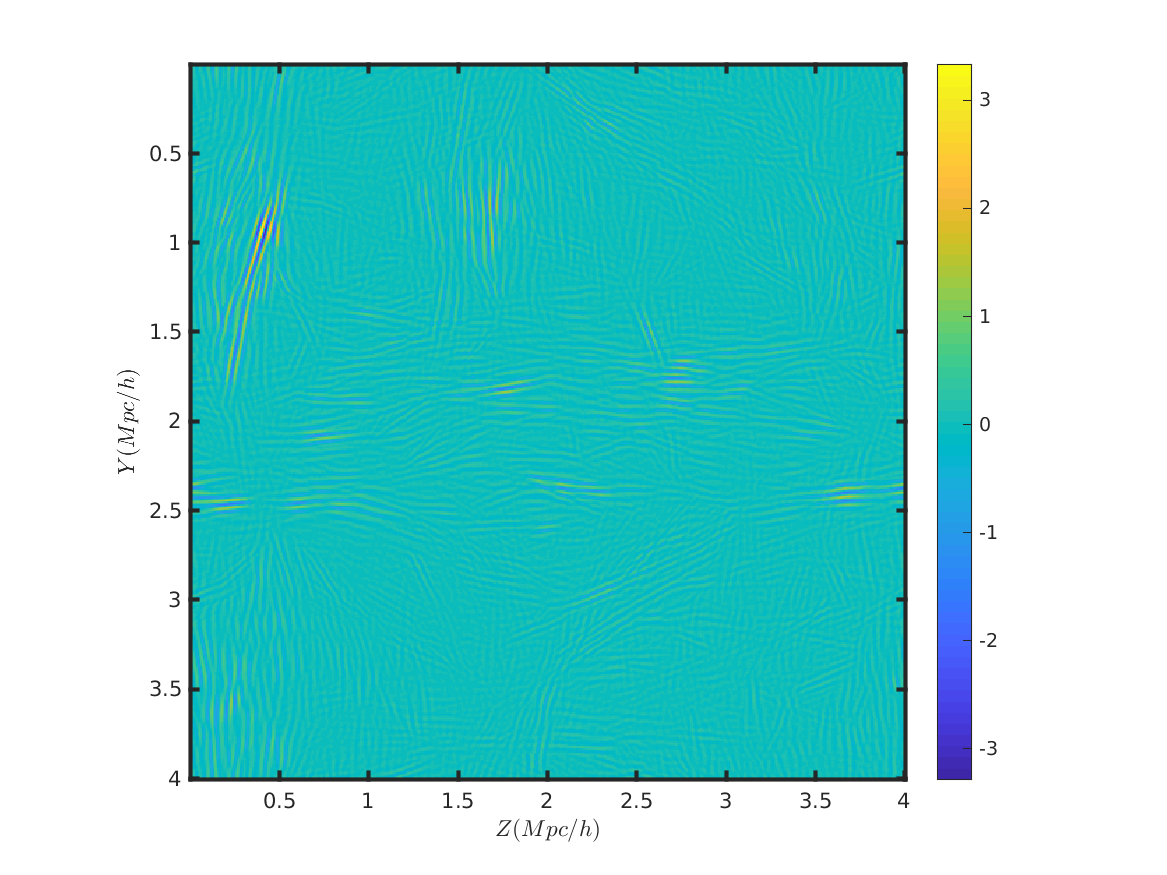}

\includegraphics[width=.45\textwidth]{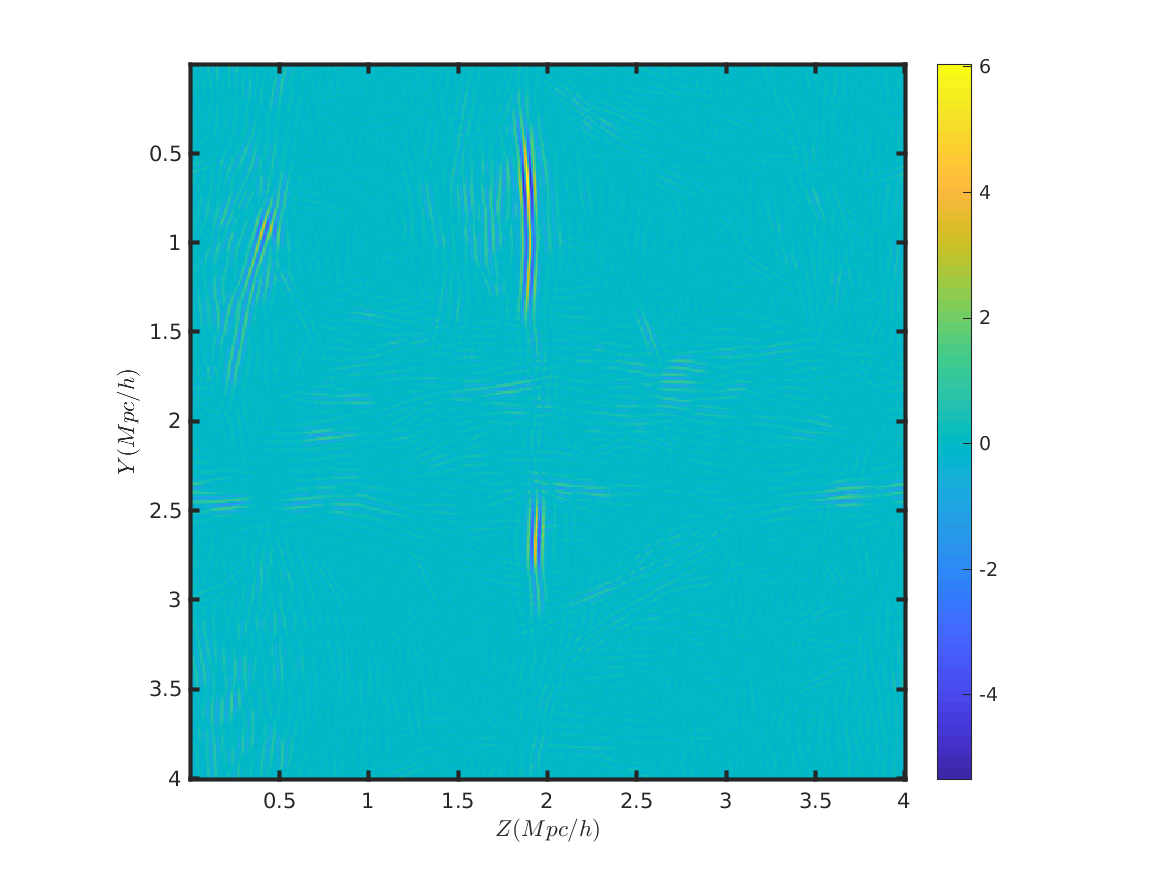}

\caption{Filtered version of a two-dimensional projection of a slice of the simulation box with lateral size $L=4Mpc/h$ and redshift $z=3$. The upper plot shows the slice without the wake, and the plot on the bottom shows the slice with a $G\mu=4 \times 10^{-8}$ wake at the center (the long vertical line at $Z\approx 2$).}
\label{2d_proj_z3s10_sl31_filter3d}

\end{figure}

Steps 3 to 6 are crucial since they highlight line segment discontinuities (as produced by the wake). To see this, consider figure \ref{2d_proj_z3s10_sl31_filter3d}, which shows the result of the sixth step associated with the maps of the figure \ref{2d_proj_z3s10_sl31}. The wake presence is manifest now, with a large line segment on the bottom panel indicating the wake position.

7-) Perform a ridgelet transformation \cite{radon} on each one of the 8 slices $rid_{k}(l,a)=R(dmFilt(i,j,k))$ , where $k$ indicates the slice, $l$ the position of the ridge, $a$ its angle . A ridgelet transformation is suitable for line detection in images, such as the ones produced by the wake. 

8-) Finally, take the wake statistical signal indicator $s(k)=pk(k)/std(k)$ of the slice $k$ as the peak $pk(k)=max(rid_{k}(l,a))$ of the ridgelet transformation divided by its standard deviation $std(k)=std(rid_{k}(l,a))$, where $max(rid_{k}(l,a))$ denotes the maximum value of the two dimensional array $rid_{k}(l,a)$ (with respect to $(l,a)$) and $std(rid_{k}(l,a))$ denotes the standard deviation value of the two dimensional array $rid_{k}(l,a)$ (with respect to $(l,a)$ also). A high $s$ value means that there is a line in the two-dimensional map which stand out very prominently with respect to the other lines, such as the one produced by the wake.

9-)  The wake presence will be indicated in the whole volume by simply summing the $s(k)$ values for each slice, producing the following wake signal $S$:

\be
S=\sum_{k=1}^{8}s(k)
\label{stat}
\ee

The analysis above is applied in two different situations: in the first one, the orientation of the wake is used as prior information, and in the second situation, this information is not used beforehand. Each case will be described in the next two subsections.

\subsection{Wake signal extraction with wake orientation prior} \label{sec_3b}
	
We have inserted the wake at the plane $Z=2Mpc/h$. Therefore, by choosing the axis $X$ as the projection axis for our analysis, we will be automatically selecting an axis perpendicular to the wake plane.

\begin{figure}[h]
\centering 

\includegraphics[width=.2\textwidth]{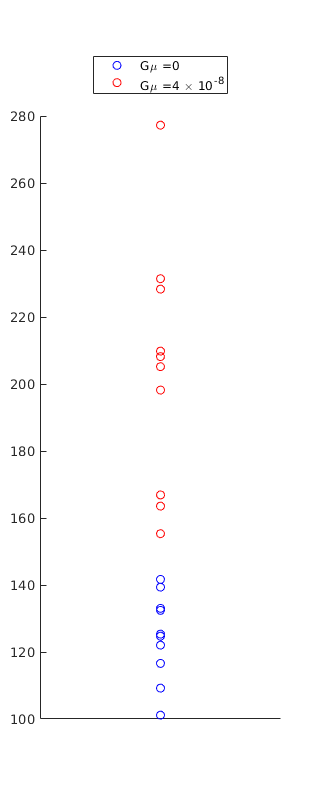}

\caption{Distribution of the wake indicator $S$ for the wake simulations (in red), and with pure $\Lambda CDM$ (in blue). }
\label{pkostd_dist}

\end{figure}

We chose the wake indicator $S$ from \ref{stat}. The result of the distribution of the signal $S$ for each one of the ten samples (with and without wakes) is shown in figure \ref{pkostd_dist}. 

In average this statistic has a confidence level of $\bar{R}=6.5$, where $\bar{R}=mean(R)$ is the mean of the signal to noise ratio for the wake simulation, defined as

\[
R=\frac{S-\bar{S}_{nw}}{std(S_{nw})}
\]
where $S$ is computed for a given simulation with a wake, $\bar{S}_{nw}$ is the mean $S$ of all simulations without wakes and $std(S_{nw})$ is the standard deviation of all $S$ from simulations without a wake. $\bar{R}$ is the mean of $R$ for all simulations with wakes.

\subsection{Wake signal extraction without wake orientation prior} \label{sec_3c}

In this subsection, we consider various orientations for the two-dimensional projections, without introducing the wake orientation information beforehand. From the observational point of view, each ${(4 Mpc/h)}^3$ volume at $z=3$ could be decomposed into a stack of 32 maps of $512 \times 512$ pixels one next to the other (in the redshift direction). To proceed with the analysis and extract as much information as we can (in this case, as many stacks of 32 two-dimensional maps), we will choose to slice the simulation cubic volume in many directions by rotating the particles and using the periodic boundary conditions, considering each direction as a different patch of the sky. It must be noted that there is a potential danger in doing that because the resulting stacks could not be independent, and assuming independence for the three-dimensional stack maps is a crucial assumption of our analysis. It is possible to ensure that the various stacks obtained by one simulation be independent by counting the degrees of freedom. At a particular snapshot, each simulation has $N_{Sim}=3\times 1024^3$ degrees of freedom (corresponding to the number of particles times three coordinates). Each stack has $N_{Stk}=32\times 512^2$ degrees of freedom (they contain 32 maps of $512^2$ pixels). Therefore we must have no more than $384=N_{Sim}/N_{Stk}$ different projection directions for constructing the stacks for each simulation. For choosing different angular orientations, we consider a Healpix set of spherical angles \citep{Gorski:2004by}, which give approximately equally-spaced adjacent spherical angles. In addition to considering different orientations, we also take advantage of the periodic boundary condition and perform random displacements and rotations on the two-dimensional figures, so the wake line does not lie on the $Z=2Mpc/h$ plane anymore. We take the parameter $N_{side}=4$ for the Healpix scheme, so we can obtain $N_{angles}=96$ different projection angles for each simulation (we are considering just non-antipodal angles since they produce equivalent two-dimensional projections), a factor of four less than what we are allowed to do.

%
%
%
%
%
%

%
%
%
%
%

We notice that there is one slice from the no-wake samples with $S>198$ and $113$ of such outliers for the simulations with wakes. In the following analysis, we will take the threshold $S_{th}=198$ as an indication of the wake presence. Therefore if in a pure $\Lambda CDM$ universe a stack set of 32 dark matter maps of $(4Mpc/h)^2$ with a resolution of $512^2$ at redshift $z=3$ is taken in the sky, the probability of it to have $S>198$ is $p_{nw}\approx 1/960=0.001$. We would naively expect that the probability of finding a similar $S>198$ dark matter volume for a universe with wakes would be $\approx 113/960=0.118$, but that is not true, since not every map of this kind will intercept a wake. With the assumption that in a $\Lambda CDM$ plus cosmic string universe there is one long string per Hubble volume, we would expect that only a fraction of about $1/24$ of the boxes \footnote{$1/24$ is the chance that a $4Mpc/h$ cube (such as the one considered in this study) will intercept one of a set of large, planar wakes about $95 Mpc/h$ apart (the Hubble radius at wake formation), corresponding one wake per comoving Hubble radius at wake formation.} would contain a wake, so we have to multiply the previous probability estimation by this factor. Therefore the expected probability of finding a similar $S>198$ dark matter map would be $p_w \approx 0.118\times (1/24)=0.0049$, which is $\approx 5$ times higher than the no wake case.

The main message from the previous paragraph is that, if we consider $10\times 96$ stacks of 32 dark matter maps with $512^2$ pixels each (which would require a field of view of $\approx 1^{\circ} \times 1^{\circ}$), we would find in average only one of those maps with $S>198$ in a universe without cosmic strings and $\approx 5$ maps in a universe with cosmic strings.  With this result now we can know what is the probability of discarding the null hypothesis that our universe is pure $\Lambda CDM$ in favor of an alternative hypothesis that our universe contains a network of $G\mu=4\times 10^{-8}$ cosmic strings. To do so, we will compute the probability of having more than $n$ outliers, given $N$ samples. Assume that the probability of a given sample to be an outlier is $p$, then the probability $P(n,N)$ of finding exact $n$ outliers among the $N$ samples is:

\be
P(n,N)=\frac{N!}{n!(N-n)!} p^n (1-p)^{N-n}
\ee

Therefore the probability $\mathcal{P}(n,N)$ of finding more than $n$ outliers among $N$ samples is:
\be
\mathcal{P}(n,N)=\sum^{N}_{m=n}P(m,N)=1-\sum^{n-1}_{m=0}P(m,N),	\label{Prob_extraOutliers}
\ee
where we make use of the fact that $\sum^{N}_{m=0}P(m,N)=1$, so the right side of \ref{Prob_extraOutliers} is much easier to compute than the middle. One can assume $p$ as $p_w$ or $p_{nw}$ computed above. For $N$ we notice that an instantaneous SKA snapshot has a field of view of $0.49\ deg^2$. This would allow for only $N\approx 600$ samples. More samples could be added by noticing that one can take extra volumes in redshifts below the ones we are considering, provided we maintain fixed the probabilities $p_w$ and $p_{nw}$ computed above and the threshold $S_{th}$. We tested that this is indeed the case for $z=2.9$, which allows us to consider $N=7680$ values\footnote{To obtain $N$ we computed the comoving volume corresponding to a line of sight depth from $z=3$ to $z=2.9$ and a field of view of ${(\Delta \theta)}^2 = 0.49\ deg^2$ and divided the result by the analysis cubic volume of ${(4Mpc/h)}^3$} for $S$, which can be taken as the number of samples for computing $\mathcal{P}$ in equation \ref{Prob_extraOutliers}. One finds that the probability $\mathcal{P}_{nw}$ of having more than $n=18$ outliers in $N$ samples of dark matter maps from a pure $\Lambda CDM$ universe is $\mathcal{P}_{nw}=0.0016$, whereas the probability of finding the same number of outliers (or more) from a universe with $G\mu=4\times 10^{-8}$ cosmic strings is $\mathcal{P}_{w}=0.9998$. Therefore if such outliers are found, we could discard a pure $\Lambda CDM$ universe with three sigmas of confidence level, and at the same time, favor a $G\mu=4\times 10^{-8}$ cosmic strings universe with three sigmas of confidence level.

%

We showed above that it is possible to identify the presence of $G\mu=4 \times 10^{-8}$  cosmic string wakes at $z\approx 3$ with a confidence level of three sigmas if a set of three-dimensional dark mater maps covering SKA-like snapshot at the range $z=3$ to $z=2.9$ is given. We would divide the entire field of view and redshift range in a set of 32 consecutive slices in the redshift direction (having $4Mpc/h$ of depth) of $(4Mpc/h)^{2}$ squares with $512$ pixels per dimension. We argued that with the most simplistic assumptions of one wake per Hubble volume, there would be more than $18$ outliers indicating the wake presence, whereas there will be no such number of outliers in the case of a universe without cosmic string wakes.

It is worth to notice that if we assume more cosmic strings per Hubble volume (numerical simulations indicate about ten cosmic strings per Hubble volume \cite{Ringeval:2010ca}) the probability $p_w$ would increase proportionally. For example, with two cosmic strings per Hubble volume one would have $p_w \approx 0.118\times (2/24)=0.0098$. The impact of this assumption for our analysis is that if 38 outliers are found, we could discard a pure $\Lambda CDM$ universe with six sigmas of confidence level, and at the same time, favor a $G\mu=4 \times 10^{-8}$ cosmic strings universe with six sigmas of confidence level.

It is also possible to extend the previous analysis (of one cosmic string per Hubble volume) by considering larger survey volumes. We considered the lowest redshift of $z=2.9$ because for lower vales, the probabilities $p_{wn}$ and $p_{w}$ do not changed. One can drop this requirement and allow lower redshifts. For example, at $z=2.5$ our simulations showed one no-wake sample with $S>200$ and $69$ samples with wakes. Therefore $S_{th}=200$ will result in $p_{nw}\approx 1/960=0.0010$ and  $p_{w} \approx 69/(24*960)=0.003$ at $z=2.5$. At higher redshifts the computation for $p_{wn}$ and $p_{w}$ for this new $S_{th}$ will change, with $p_{wn}$ gradually decreasing and $p_{w}$ gradually increasing. This means that for the full range $z=3$ to $z=2.5$ one can assume $p_{nw} \leq 0.0010$ and  $p_{w} \geq 0.003$. Although the probability of having detecting a wake is smaller than the previous case, the survey volume will be about seven times higher, which mean we can take $N=5 \times 10^{4}$. By computing $\mathcal{P}_{nw}$ and $\mathcal{P}_{w}$, one arrive at the conclusion that the probability of finding $93$ outliers with $S>200$ given the $N=50000$ samples provided by a SKA snapshot in the range $z=3$ to $z=2.5$ is $\mathcal{P}_{nw} \lesssim 2.0 \times 10^{-7}  $ and $\mathcal{P}_{w} \gtrsim 1-2.5\times 10^{-7}$, allowing to favor a cosmic string universe over of a pure $\Lambda CDM$ universe with five sigma of confidence level.

One can extend even further and consider not only the SKA instantaneous snapshot in the range $z \in [3,2.5]$, which spans $\Delta \theta = 0.49\ deg^2$, but its full scan of a solid angle about $\approx 9900$ larger, and therefore $N\approx 4.5\times 10^{8}$. To handle such a high number, one used the MPFUN2015 \cite{David:2017} Fortran-90 multi-precision package, that allows high-precision computations for large numbers. One finds that the probability of finding more than $53600$ outliers is $\mathcal{P}_{nw} \lesssim 1.0 \times 10^{-66}$ and $\mathcal{P}_{w} \gtrsim 1-\times 10^{-66}$, meaning that there is a potential to exclude a pure $\Lambda CDM$ universe in favor of a universe with cosmic strings with more than seventeen sigmas of confidence level.




\section{Conclusion}

With this work, we can affirm that a sky map of the dark matter distribution can be used do constrain the existence of cosmic strings with high statistical significance. 

We are investigating whether neural networks could improve wake detection. The hope is that they will be able to distinguish maps with and without wakes at redshifts below the range $z \in [3,2.5]$ and lower string tension parameter. Regarding the first extension, the main challenge will be computational since the non-linearities cause the simulation runs to suffer a substantial increase in time. As for the lower string tension, the challenge will also be computational, since at $G\mu \leq 4\times 10^{-8}$, the wake becomes not well resolved at the time of insertion, causing the simulations not to accurately capture the dynamics at small scales required to evolve the fainter wake.  A preliminary analysis was made for the $G\mu = 2\times 10^{-8}$ wake, and we found a confidence level of $\bar{R}=1.8$ for the analysis of section III-b (with orientation prior) and a probability of finding a wake equal to the probability of not finding one for the analysis of sections III-c  (without orientations prior). Therefore the analysis developed in this work is not able to extract a $G\mu = 2\times 10^{-8}$ wake signal, confirming the expectation that simulations with more resolutions should be made in order to extract the signal of fainter wakes. A possible way to circumvent the computational bottleneck imposed by these challenges is to use COmoving Lagrangian Acceleration (COLA) methods \cite{Tassev:2013pn}, but consistency check must be performed in order to certify that the small scales are adequately represented. Experimentally, it will also be challenging to extract a fainter wake signal, since the SKA-mid angular resolution used in this work will not be enough to resolve it.


The experimental prospects to find the wake signal will be analyzed in future works, where resolution (angular and redshift) is essential together with intensity sensitivity. Finally, it remains to be seen if the dark matter tracers, such as halos and galaxies, will maintain the wake signal. All of those aspects are under current investigation.

\section*{Acknowledgement}
\noindent

The author is grateful to Dr. Robert Brandenberger, Dr. Adam Amara and Dr. Joachim Harnois Deraps for useful discussions and feedback, and also would like to thank the anonymous referee for his/her valuable suggestions on this work. He also wishes to acknowledge CAPES (Science Without Borders) for a student fellowship, NSERC Discovery Grant, which supported the research at McGill and FNRS, which made possible the research at CURL. This research was enabled by the support provided by Calcul Quebec (http://www.calculquebec.ca/en/), SciNet (https://www.scinethpc.ca) and Compute Canada (www.computecanada.ca).


\end{document}